\documentclass[sigconf, nonacm]{acmart}

\usepackage{xspace}
\usepackage{paralist}
\usepackage[ngerman,english]{babel}
\usepackage{hyperref}
\usepackage{enumitem}
\usepackage{balance}

\makeatletter
\newcommand{\customlabel}[2]{#2\def\@currentlabel{#2}\label{#1}}
\makeatother

\newcommand{\degreeOfPollution}{degree of pollution\xspace}
\newcommand{\inKommaSatz}[1]{\glqq{#1}\grqq}
\newcommand{\phasetitle}[1]{\textbf{#1}}
\newcommand{\Toolname}{DaPo$^+$\xspace}
\newcommand{\ToolnameOld}{DaPo\xspace}

\begin{document}
\title[Towards Scalable Generation of Realistic Test Data for Duplicate Detection]{Towards Scalable Generation of Realistic Test Data \\ for Duplicate Detection}

\author{Fabian Panse}
\affiliation{%
  \institution{Hasso Plattner Institute}
  \city{Hamburg}
  \country{Germany}
}
\email{fabian.panse@hpi.de}

\author{Wolfram Wingerath}
\affiliation{%
  \institution{University of Oldenburg}
  \city{Oldenburg}
  \country{Germany}
}
\email{wolfram.wingerath@uol.de}

\author{Benjamin Wollmer}
\affiliation{%
  \institution{University of Hamburg}
  \city{Hamburg}
  \country{Germany}
}
\email{benjamin.wollmer@uni-hamburg.de}

\begin{abstract}

Due to the increasing volume, volatility, and diversity of data in virtually all areas of our lives, 
the ability to detect duplicates in 
potentially linked data sources 
is more important than ever before.
However, while research is already intensively engaged in adapting duplicate detection algorithms to the changing circumstances,
existing test data generators are still designed for small -- mostly relational -- datasets 
and can thus fulfill their intended task only to a limited extent.
In this report, 
we present our ongoing research on a 
novel approach for test data generation
that -- in contrast to existing solutions --
is able to produce large test datasets 
with complex schemas and more realistic error patterns
while being easy to use for inexperienced users.

\end{abstract}

\maketitle

\section{Introduction}
\label{Introduction}

The detection of duplicate records is a critical task in data cleaning~\cite{RahmD00,GantiS13,IlyasC19},
data integration~\cite{Doan2012,DongS2014}, and data linkage~\cite{Christen2012Book}.
It has been extensively studied during the last decades~\cite{Nau2010,Talburt12,2021Papadakis}
and is still subject of many research projects today~\cite{KushagraSIB19,0001LSDT20,FrankeSRR21,MudgalLRDPKDAR18,GazzarriH21,WuCSCT20,KoumarelasPN20}.

Quality evaluation using appropriate test data is an essential aspect 
in developing new duplicate detection algorithms
and adjusting existing ones to specific use cases.
The most adaptable and flexible approach 
is to use a test data generator,
such as DBGen~\cite{HernandezS98}, GeCo~\cite{ChristenV13}, or EMBench$^{++}$~\cite{IoannouV19}.
The goal of these generators is to create test data that
resembles the required real-world properties 
(e.g., sparsity or textuality~\cite{PrimpeliB20}) as well as possible.

However, according to Gardners' 4 \inKommaSatz{V}s, 
datasets are becoming bigger (volume), more heterogeneous (variety),
and are changing faster (velocity) than ever before
while often being incomplete and error-prone (veracity).
Thus, while the ever-increasing number of data sources makes
the integration of data more and more valuable,
the size, velocity, diversity, and error-proneness of the individual sources 
make this task more and more challenging~\cite{2021Papadakis}.
In conclusion, duplicate detection will remain an important task in data management.
But while there is already plenty of research 
on developing new algorithms adapted to these circumstances (e.g.,~\cite{DuttaNB13,SaeediNPR18,MaLWCP19}),
there is little to no work addressing
the generation of appropriate test data 
(cf. Section~\ref{Related Work}).
Therefore, we strive in our research for
a novel approach for test data generation
that in contrast to existing solutions
\begin{itemize}[topsep=1pt]
\item[\textbf{\customlabel{contribution:scalability}{c\textsubscript{1}}:}]
scales vertically as well as horizontally and thus well enough to generate datasets with millions of records,
\item[\textbf{\customlabel{contribution:datamodels}{c\textsubscript{2}}:}]
supports NoSQL data models (e.g., JSON or property graphs),
\item[\textbf{\customlabel{contribution:configuration}{c\textsubscript{3}}:}]
provides an automatic preconfiguration to aid novice users, 
\item[\textbf{\customlabel{contribution:complex}{c\textsubscript{4}}:}]
supports the generation of complex scenarios with multiple data sources,
each defined on a complex data schema, and
\item[\textbf{\customlabel{contribution:errormodel}{c\textsubscript{5}}:}]
utilizes an event-based error model 
to realistically simulate temporal errors and complex error patterns
as they result, e.g., from outdated values and copying processes.
\end{itemize}

The remainder of this paper is structured as follows. 
First, we describe three different contexts with their individual characteristics in which duplicate detection is applied.
Then, we discuss existing test data generators and their limitations in Section~\ref{Related Work}.
Thereafter, we present the idea of our new generation approach  in Section~\ref{DaPo} and discuss challenges as well as our ongoing research in Section~\ref{Open Challenges}.
Finally, we conclude the paper in Section~\ref{Conclusion}.

\section{Application Contexts}
\label{Application Contexts}
Since the detection and elimination of duplicates is needed in various contexts,
the requirements for corresponding algorithms can be very different.
Roughly speaking, these contexts can be divided into three categories:
\begin{itemize}[leftmargin=.5cm,topsep=2.0pt]
\item \textbf{Data Cleaning:} 
If we want to clean a single data source from intra-source duplicates,
all records conform to the same schema 
and most of them are defined using 
the same formats, units and language.
Thus, the degree of heterogeneity is rather low
and divergences between duplicate records primarily result from data errors and outdated values.
\item \textbf{Data Integration:}
If we want to remove inter-source duplicates from an integration result, 
all records conform to the same (target) schema,
but many of them are defined in different formats, languages, vocabularies and units of measurements.
Thus, the degree of heterogeneity is much higher than in a data cleaning scenario.
Depending on the quality of the integration process
this can also include differences on the schema level
(e.g., values have been mapped to wrong attributes).
\item \textbf{Data Linkage:}
If we want to link inter-source duplicates that have not been integrated into a common target schema so far,
these records are not only defined in different formats, units, etc.,
but also conform to different schemas.
Thus, the given records are much more heterogeneous than in the previous two categories.
\end{itemize}

In the case of duplicate detection, the gold standard corresponds to a duplicate clustering (one cluster per real-world entity)~\cite{MenestrinaWG10}.
In the case of duplicate elimination, 
we additionally need the correct values for every cluster, 
the so-called \emph{golden record}~\cite{Deng2019}.

\section{State of the Art \& Related Work}
\label{Related Work}

Static test datasets (e.g., Magellan project~\cite{DoanKCGPCMC20})
are very useful when it comes to getting a rough impression of an algorithm's quality,
but cannot be used to systematically evaluate the algorithm's behavior with respect to 
changing conditions such as the data's volume, quality, or complexity.
Instead, this requires a test data generator 
that allows to control the number of records, duplicates, and data errors.
Test data generators can be divided into two classes.
\emph{Data synthesis} tools create a whole dataset including duplicates and errors from scratch.
In contrast, \emph{data pollution} tools 
inject duplicates, errors, and heterogeneities into an already existing dataset.

\vspace*{0.1cm}

\noindent\phasetitle{Existing Tools.}
Existing data synthesis tools, 
such as DBGen~\cite{HernandezS98} or the Febrl dataset Generator~\cite{Christen09},  
are very efficient so that large datasets can be generated in short time.
However, since all values are fictional
and based on handmade generation rules,
they struggle to generate realistic data
and are limited to a single domain.
To overcome this drawback, 
current approaches for general-purpose data synthesis~\cite{PingSH17,2020ElEmam}
(data without duplicates and errors)
use Deep-Learning-techniques (e.g., GANs), 
but this comes with longer runtimes.
Data pollution tools,
such as DirtyXML~\cite{Puhlmann04}, GeCo~\cite{ChristenV13}, TDGen~\cite{Bachteler12}, 
EMBench$^{++}$~\cite{IoannouV19}, 
BART~\cite{ArocenaGMMPS15}, or Lance~\cite{SavetaDFFN15},
are designed to generate test data with realistic value patterns
because real-world data can be used as input.
Moreover, if a broad spectrum of error classes is supported, they are nearly domain-independent.

\vspace*{0.1cm}

\noindent\phasetitle{Limitations.}
Existing generators have several limitations:
\begin{itemize}[leftmargin=.5cm,topsep=2.0pt]
\item \textbf{Data Size (cf. \ref{contribution:scalability}):}
Existing data pollution tools are strongly limited with respect to their scalability (runtime as well as memory)
making a generation of large datasets either impossible or too time-consuming~\cite{Hildebrandt2020}.
However, in times of big data, many algorithms focus on scalability 
(e.g., \cite{RastogiDG11,GetoorM13,SaeediNPR18,ChenSS18})
so that an evaluation of their key functionalities requires large test 
datasets with millions or even billions of records.

\item \textbf{Model Diversity (cf. \ref{contribution:datamodels}):}
They are limited to relational (e.g., DBGen, TDGen), XML (e.g., DirtyXML), and/or RDF data (e.g., Lance) 
although the database landscape became extremely diverse in recent years
and different kinds of NoSQL systems, 
such as document stores or graph databases
find increased use in practice~\cite{GessertWFR17}.
In addition, a matching of records across individual data models
is extremely relevant if we want to integrate or link data in highly diverse environments.

\item \textbf{Number of Sources (cf. \ref{contribution:complex}):}
Most of them generate a single dataset, 
although the evaluation of duplicate detection algorithms
for data integration and data linkage requires the existence 
of multiple data sources having different schemas.

\item \textbf{Schema Complexity (cf. \ref{contribution:complex}):}
Most of them generate a single table, 
although in the last decades a lot of research
(e.g., \cite{DongHMN05,BhattacharyaG07,RastogiDG11,BronselaerBT15,Efthymiou0SC19}) 
focused on a detection of duplicates in complex datasets 
where records are connected via different kinds of relationships (e.g., modeled by using foreign keys or RDF triples).
To the best of our knowledge, there is currently no study
which compares the quality of these algorithms
with respect to datasets of varying schema complexity.

\item \textbf{Automatic Configuration (cf. \ref{contribution:configuration}):}
Almost none of them provides any mechanism to 
compute parameter settings automatically,
although it is essential
to enable a proper usage of the generator by inexperienced users.

\item \textbf{Copying Patterns (cf. \ref{contribution:errormodel}):}
There are several works (e.g., \cite{DongBHS10,PochampallySDMS14})
that use copying relationships between data sources to increase the quality of their duplicate elimination results.
Existing generators, however, inject errors into datasets by using simple mechanisms
and thus do not support the generation of complex error patterns
as they result from those copying processes.

\item \textbf{Outdated Values (cf. \ref{contribution:errormodel}):}
With increasing velocity,
outdated data become a major quality issue \cite{ZhengMC19}.
However, existing generators are very bad 
at simulating a consistent appearance of outdated values across individual records or data sources.

\end{itemize}

\section{Test Data Generation in Six Phases}
\label{DaPo}

In our previous work~\cite{Hildebrandt2020},
we proposed the data polluter \ToolnameOld 
that scales horizontally by using Apache Spark.
The implementation of this prototype, however, focused on the scalability aspect
and did not address the other shortcomings listed in Section~\ref{Related Work}.
With \Toolname, we are extending this prototype 
to support non-relational data models, 
more complex data schemas, and more realistic error patterns.

\begin{figure}[t]
	\centering
		\includegraphics[scale=0.57]{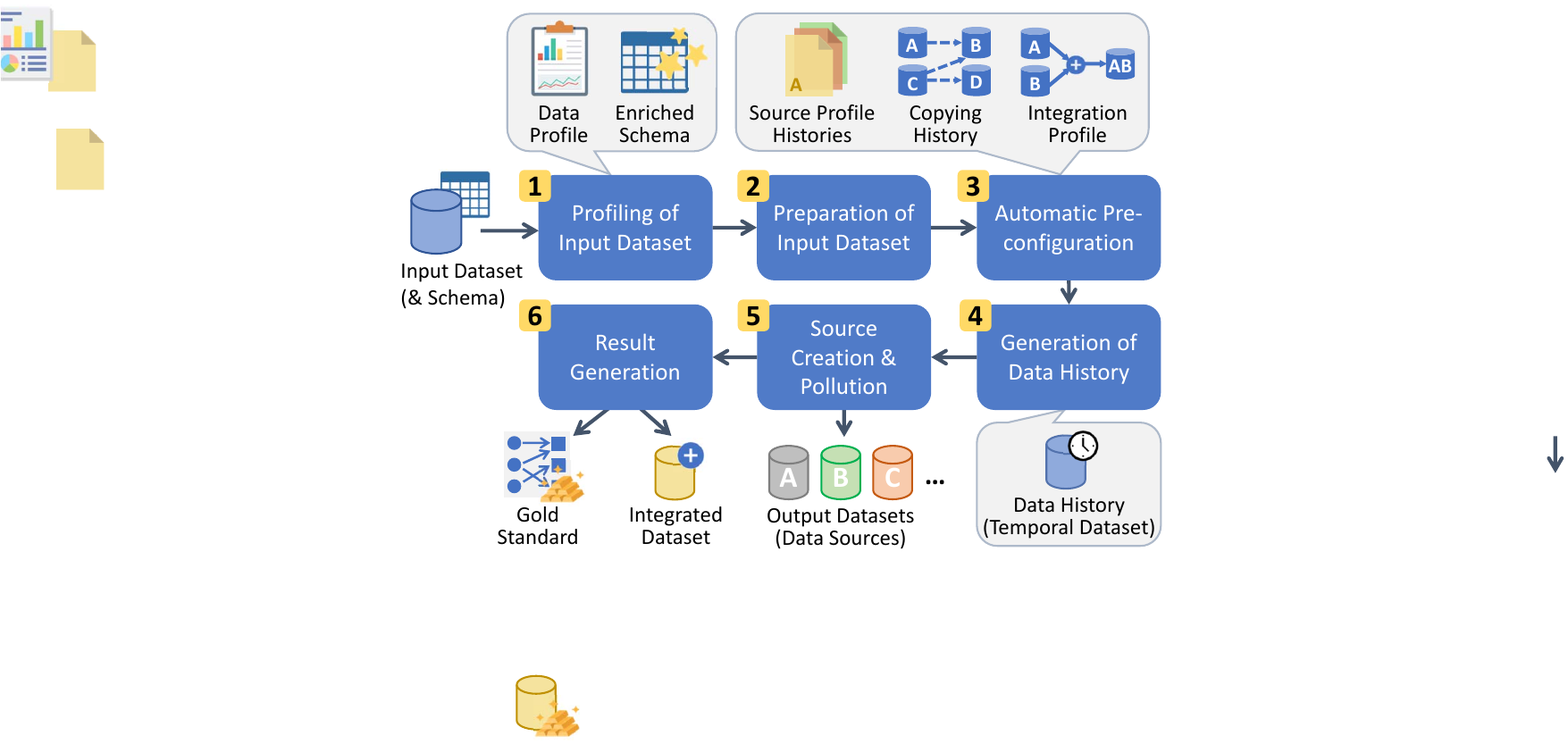}
		\caption{Basic architecture of \Toolname}
	\label{fig:WorkflowSecondPrototyp}
\end{figure}

As shown in Figure~\ref{fig:WorkflowSecondPrototyp},
\Toolname gets a single clean dataset (real-world or synthesized) as input 
and produces either a \emph{cleaning}, \emph{integration}, or \emph{linkage scenario} as output.
In the cleaning scenario, we generate a single dataset with intra-source duplicates.
In the latter two scenarios, we generate a predefined number of datasets
with intra- and inter-source duplicates and integrate these datasets into a single
target schema in case of the integration scenario.
All three scenarios also contain 
a so-called \emph{gold standard} modeling the true duplicate relationships between their records.
To properly reflect the variety, veracity, and velocity of many real-world use cases,
the output datasets contain a user-controlled degree of data errors and may vary in their schemas.
Since these datasets serve as \emph{data sources} 
in the individual scenarios,
we also refer to them as such.

The architecture of \Toolname includes six phases (see Figure~\ref{fig:WorkflowSecondPrototyp}).
In the first four phases, the input dataset is analyzed, prepared, and enriched.
In addition, a proper configuration of the actual generation process 
is calculated based on the user's specifications.
In the final two phases, 
the actual test data is generated.

\vspace*{0.1cm}

\noindent\phasetitle{1. Profiling.} 
First, the input dataset is analyzed.
The results are a data profile and an enriched data schema.
The profile contains statistics about 
the individual attributes (e.g., length and number of tokens)
and relationships between records (e.g., frequency and type).
The enriched schema contains important metadata such as
integrity constraints (e.g., functional dependencies),
semantic column types, and
temporal characteristics (e.g., update frequencies and dependencies).

\vspace*{0.1cm}

\noindent\phasetitle{2. Preparation of Input Dataset.}  
The transformation of the input data to the data models and schemas of the generated data sources 
as well as the injection of errors are usually much easier to handle
if these sources are modeled on a lower level of detail than the input dataset
(e.g., it is easier to merge two attributes than to split one).
Therefore, the profiling results 
are used to prepare the input dataset and the enriched data schema.
Among others, this includes transforming it into a structured data model, 
normalizing its schema, and splitting its attributes into several subattributes 
if a clear separation between the corresponding values is possible.

\vspace*{0.1cm}

\noindent\phasetitle{3. Automatic Preconfiguration.} 
Since \Toolname is a complex system and one of our goals is to support inexperienced users,
a preconfiguration is automatically derived from the enriched data schema.
This configuration contains a 
\emph{source profile history} (i.e., a sequence of \emph{source profiles} where every profile has a validity period)
for each of the data sources that have to be created in the current generation task.
Each of the source profiles in turn consists of a 
\emph{representation profile} and an \emph{error profile}.
A representation profile determines in which way which data is modeled within the described source.
This includes the source's data model and schema as well as its scope (i.e., the relevant parts of the real-world).
An error profile determines in which way the source's data is corrupted by errors,
including the amount of duplicates and the probabilities of certain error classes.
In addition to the source profile histories, 
a \emph{copying history} is generated.
It determines which sources copy which data from which other sources during which period of time.
This includes a program that transforms data from the copied to the copying source (including possible transformation errors).
In case of an integration scenario, an additional integration process is required
whose configuration is called \emph{integration profile} in Figure~\ref{fig:WorkflowSecondPrototyp}.

\vspace*{0.1cm}

\noindent\phasetitle{4. Generation of Data History.} 
To enable the simulation of outdated values and data copying as defined in the corresponding configuration files,
a data history is generated 
based on the given input data, the data profile, and the enriched data schema.

\vspace*{0.1cm}

\noindent\phasetitle{5. Source Creation \& Pollution.}  
The required number of data sources is created
based on the data history, the source profile histories, and the copying history. 
To create these sources and inject duplicates as well as errors into their data,
we use our novel event-based error model (see Section~\ref{Data Pollution}).

\vspace*{0.1cm}

\noindent\phasetitle{6. Result Generation.}
The final output (including the gold standard)
is generated based on the previously generated data sources.
If an integration scenario is requested, 
these sources are integrated into a single target schema
by using the integration profile. 
To increase the value of the costly generated data sources without much additional effort,
we use more than one target schema 
to generate several integration scenarios in a single run.

\section{Challenges \& Ongoing Research}
\label{Open Challenges}

In this section, we describe the various challenges we face in our project
and present the current status of our research in the development of suitable solutions.
Since our current prototype has already proven to scale very well~\cite{Hildebrandt2020},
we limit the following discussions to the remaining challenges.

\begin{figure}[t]
\centering
    		\includegraphics[scale=0.38]{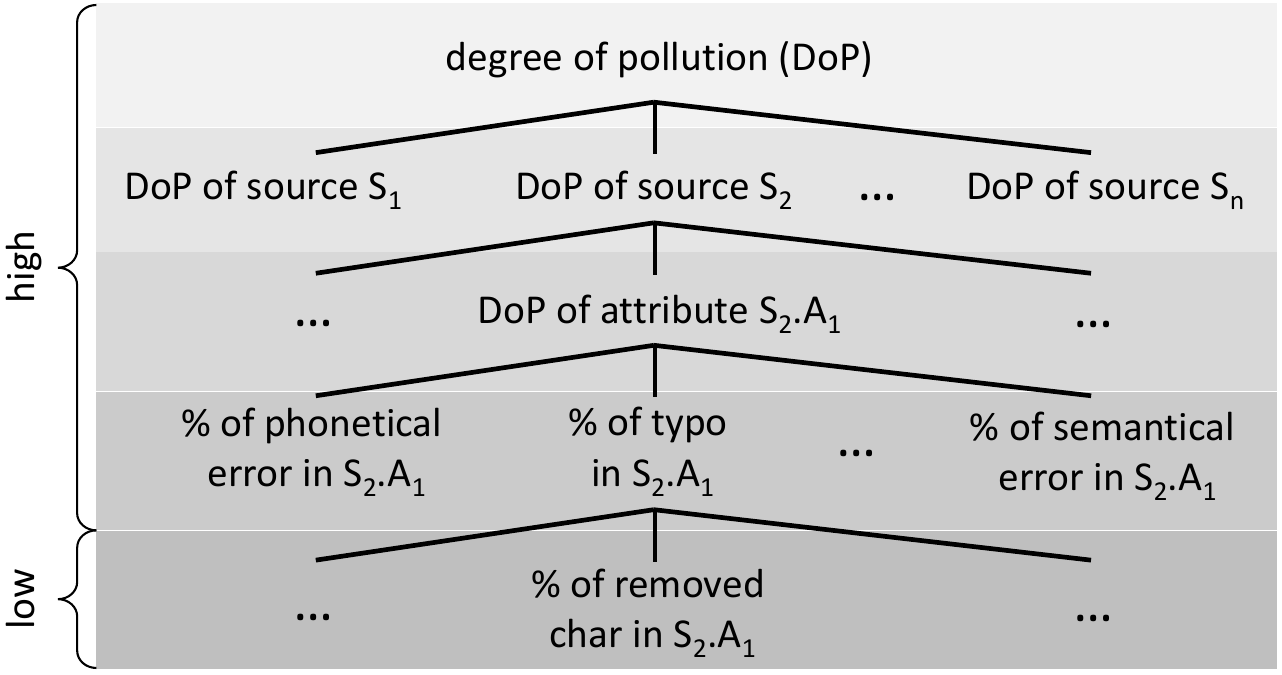}
			\caption{Sample hierarchy of error parameters. 
			Automated mappings from the few abstract high-level parameters 
			to the many low-level parameters
			support inexperienced users in configuring the actual error probabilities.}
			\label{fig:ParameterHierarchy2}	
\end{figure}

\subsection{Data Profiling}
\label{Data Profiling}

The more metadata we have about the input dataset, 
the more realistic are the data history, the schemas as well as the errors generated 
in the subsequent phases~\cite{PanseN21}.
Therefore, we are developing a data profiling component which is able 
to extract useful metadata from the given input data.
In case of NoSQL input data even whole schema versions may have to be extracted~\cite{KlettkeSS15}.

As surveyed by Abedjan et al.~\cite{AbedjanGN15},
there is already a lot of research on collecting database statistics,
identifying semantic domains~\cite{HulsebosHBZSKDH19,ZhangSLHDT20}, 
and detecting dependencies
(exact, conditional, or approximate) \cite{BirnickBFNPS20,PenaAN19,LivshitsHIK20,TschirschnitzPN17,Berti-EquilleHN19,SchirmerP0NHMN19,CaruccioDNP21}.
Most of these projects, however, focus on relational data
and only a few of them consider XML \cite{VincentLL04}, RDF \cite{AbedjanGJN14}, or JSON \cite{MollerBKSS19,Mior2021} data.
In addition, only few projects~\cite{AbedjanAOPS15,BleifussBJKNS18,ZhengMC19} 
address the detection of temporal data characteristics. 
However, to generate realistic data histories, we need to identify:
\begin{itemize}[leftmargin=.5cm,topsep=2.0pt]
\item \textbf{Changes:} 
when, how often, and how (what are allowed and typical updates?) records change, 
\item \textbf{Dependencies:} 
intra-record (e.g., phone landline number $\leftrightarrow$ residence)
and inter-record (e.g., residence of family members or pay raise of co-workers) dependencies, and 
\item \textbf{Integrity:} 
constraints that are not allowed to be violated at any point in time
(e.g., a key has always to be unique)
or over all points in time (e.g., an old key value is never reused).
\end{itemize}

Our main focus is on the adaptation of existing profiling algorithms to non-relational data models
and the development of profiling algorithms to detect temporal data characteristics. 
For instance, intra-record dependencies can be modeled as association rules,
e.g., $\textit{update of }A \rightarrow \textit{update of }B$.
Thus, our current solution to identify 
those dependencies is to leverage existing techniques for frequent-itemset mining (FIM)~\cite{TanSK2005}.
The identification, however, is more difficult 
when there is a small time lag between the two updates
so that they are assigned to different FIM-transactions. 
To address this problem, 
we calculate support values by
using a sliding window over the time-ordered list of transactions
(i.e., all transactions within the same window are treated as one).

\subsection{Automatic Preconfiguration}
\label{Automatic Preconfiguration}

\Toolname's large number of parameters will allow users
to configure the generation process in a flexible and detailed way
and thus helps them to generate test data that meet their requirements best.
Setting these parameters to proper values, however, requires a lot of domain knowledge 
and experience in using this generator. 

Since it is almost impossible to reduce the number of parameters without reducing functionality,
we intend to relieve users by introducing a few abstract high-level parameters 
which are set by them and then serve as a basis 
to calculate proper settings for the actual (low-level) parameters by the system itself
(before and/or during runtime).
To allow experienced users to invest more effort in the configuration process, 
we introduce a hierarchy of abstract high-level parameters
so that they can enter the configuration at a granularity level of their choice.
We illustrate such a potential hierarchy for the \degreeOfPollution in Figure~\ref{fig:ParameterHierarchy2}
where a single \degreeOfPollution for the whole dataset is at the top
and the probabilities of the actual error classes are at the bottom.

Our current efforts are aimed at identifying
\begin{itemize}[leftmargin=.5cm,topsep=2.0pt]
\item \textbf{High-level Parameters} 
that are intuitive, simple to understand, and easy to map to the actual low-level parameters.
\item \textbf{Efficient Measures} 
to calculate high-level parameters (e.g., the current degree of pollution),
so that we are able to adapt the corresponding low-level parameters dynamically at runtime. 
\end{itemize}

A special challenge is the reasoning of the low-level parameters 
that address the representation profile (data model, schema, etc.) of the individual data sources.
Here we are reusing concepts of existing schema generators, such as iBench~\cite{ArocenaGCM15}.

\subsection{Generation and Reuse of Data Histories}
\label{Data History}

A major goal of \Toolname is to enable the generation of realistic outdated values and the simulation of copying processes between data sources
by using a data history. 
This history is either part of the input or needs to be generated based on the profiling results.

\vspace*{0.1cm}

\noindent\phasetitle{Generation \& Realism.} 
We aim to generate a data history by modifying the given input dataset,
which is either a single snapshot or already a data history.
This includes the update or deletion of existing records and the insertion of new records.
In complex data schemas, such modifications also address relationships between records.
To make these modifications as realistic as possible
every intermediate state of the generated data history
has to comply with the integrity constraints of the enriched data schema.
This poses a major challenge to the scalability of the generation process
if we consider constraints, such as uniqueness, that take large parts of the dataset into account.
To further increase the realism of the resulting history, 
we build on the latest achievements in 
data synthesis~\cite{2020ElEmam}
and extend them by a proper handling of temporal aspects.

\vspace*{0.1cm}

\noindent\phasetitle{Reusability.} 
If the input dataset comes without any history,
we want to adopt temporal data characteristics from a similar dataset
because having non-perfect training data is often still better than having no training data at all.
Therefore, we maintain a repository of datasets 
and the update models (e.g., a set of rules) learned from them. 
This requires appropriate measures to capture similarities between datasets
(e.g., based on summaries as they are calculated by DataSynthesizer~\cite{PingSH17})
and methods to adapt update models from one dataset to another.

\begin{figure}[t]
  \centering
  	\includegraphics[scale=0.435]{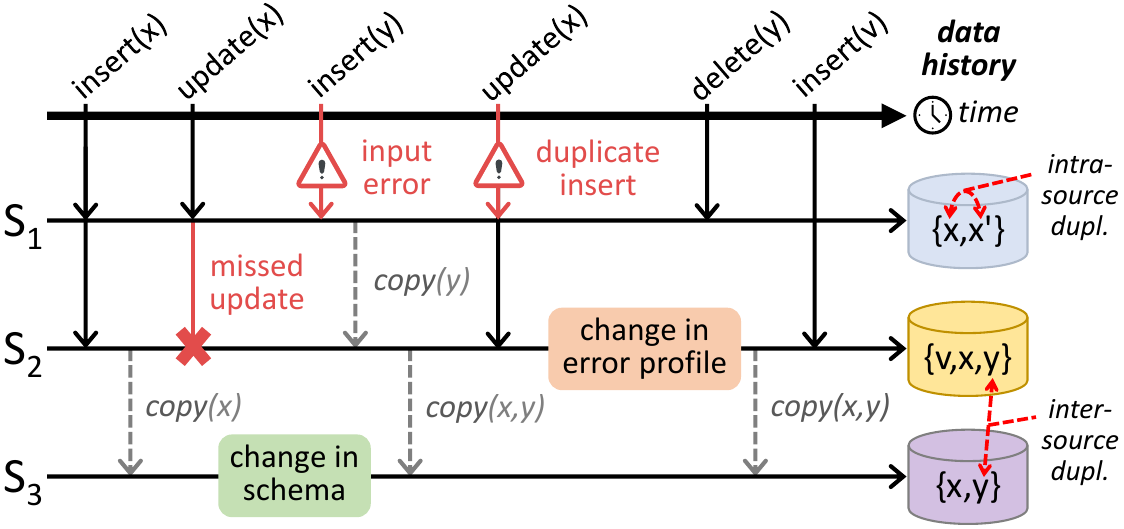}
	\caption{Event-based error model with three data sources.
	Events are inserts, updates, and deletes of records, 
	copying processes, and changes in the schema and error profiles.}
	\label{fig:Event-basedModel}
	
\end{figure}

\subsection{Source Creation \& Pollution}
\label{Data Pollution}

The error model defines in which way 
the different histories (data, source profile, and copying) are used
to generate the individual sources 
and to pollute their data.
The error models typically used by data pollution tools (including \ToolnameOld) 
take the input dataset as basis and then subsequently replace single values (or whole records)
with copies modified by using a specific error class~\cite{IoannouV19}.
If we stick to this approach,
we can simulate outdated values in a consistent fashion
by using the data history as a kind of look-up table.
Every time we want to set a value to an old instance,
we compute a past point in time randomly 
and then replace the value's current instance with the corresponding instance from the data history.

These error models can be executed efficiently
because they pollute most parts of the dataset independently.
However, they have several limitations when it comes to temporal aspects.
One of them concerns the integration of time-variant copying patterns
because all data records are treated individually
so that a time-based consistency across these records is hard to achieve.
Similar problems apply to temporally triggered changes within 
the error or representation profiles,
such as an evolving data schema.

\vspace*{0.1cm}

\noindent\phasetitle{Event-Based Error Model.} 
To address these problems,
we developed an event-based error model (see Figure~\ref{fig:Event-basedModel}).
The basic idea of this model
is to simulate the whole time-span from the beginning to the end of the data history 
by processing events, 
which originate from the different input histories
and the reactions of the data sources to these events.
Those events include:
\begin{itemize}[leftmargin=.5cm,topsep=2.0pt]
\item changes of real-world conditions (modeled by the data history) 
provoking inserts, updates, and deletes in the data sources, 
\item copying processes triggered either periodically or by an insert, update, or delete in one of the data sources, and
\item changes of the representation profile (e.g., data model, schema, or scope)
or error profile (errors are fixed or new ones arise) of an individual data source.
\end{itemize}
By doing so, the copying pattern can be incorporated easily by considering every copying process as an event.
Outdated values are simply created when a data source misses to update its data according to a changing real-world condition.
Errors can be introduced into the data in different ways and at a very fine level of granularity.
Typos, formatting, or phonetic errors are immediately inserted into the values when a data source
reacts to an event (e.g., insert, update, or copying)
or can be inserted afterwards to simulate errors in data maintenance.
Dependencies between errors, as they result when different values/records are affected by
the same faulty hardware (e.g., an infrared sensor) or software (e.g., an ETL library) component, 
can be modeled and incorporated into the data pollution process very smoothly.
The same holds for time-variant changes of such error sources 
(e.g., it can be simulated that after some time a data source detected a damaged sensor and replaced it).
In summary, the event-based error model allows 
to model the different aspects of a dynamic world 
in a very detailed manner.

\vspace*{0.1cm}

\noindent\phasetitle{Error Classes \& Patterns.} 
Due to the structural differences of NoSQL data models and the different behavior of NoSQL data stores, 
NoSQL datasets can contain other types of errors than relational ones 
(e.g., an incorrectly sorted list in a JSON document).
Therefore, we are developing new error classes and mechanisms to integrate them into the error model.
An integration into the event-based error model is quite simple,
because we are able to mimic typical processes 
that lead to errors in real life one-to-one.

\section{Conclusion}
\label{Conclusion}
In both research and practice, benchmarking is an important instrument to ensure algorithms of high quality.
In this paper, we discussed the limitations of existing test data generators for duplicate detection
including the lack of support for complex error patterns, temporal errors, and inexperienced users.
We then presented our approach for a novel test data generator called \Toolname
and described how it addresses these issues.
This approach introduces a new set of challenges, such as the generation of a data history, 
which we described along with our ongoing research to solve them.

\balance
\bibliographystyle{ACM-Reference-Format}
\bibliography{literaturDB2}

\end{document}